\documentclass[a4paper,onecolumn,11pt]{quantumarticle}
\pdfoutput=1
\usepackage[utf8]{inputenc}
\usepackage[english]{babel}
\usepackage[T1]{fontenc}
\usepackage{amsmath,amssymb,amsfonts}
\usepackage{mathtools}
\usepackage{graphicx}
\usepackage{hyperref}
\usepackage[braket,qm]{qcircuit}

\hypersetup{
    colorlinks=true,
    linkcolor=blue,
    citecolor=blue,
    filecolor=blue,
    urlcolor=blue,
}

\newcommand{\multiprepareC}[2]{*+<1em,.9em>{\hphantom{#2}}\save[0,0].[#1,0];p\save !C
  *{#2},p+RU+<0em,0em>;+LU+<+.8em,0em> **\dir{-}\restore\save +RD;+RU **\dir{-}\restore\save
  +RD;+LD+<.8em,0em> **\dir{-} \restore\save +LD+<0em,.8em>;+LU-<0em,.8em> **\dir{-} \restore \POS
  !UL*!UL{\cir<.9em>{u_r}};!DL*!DL{\cir<.9em>{l_u}}\restore}

\title{Robust decompositions of quantum states}
\author{Jonathan E. Moussa}
\email{godotalgorithm@gmail.com}
\affiliation{Molecular Sciences Software Institute, Virginia Tech, Blacksburg, VA 24060}
\date{}
\begin{document}
\maketitle

\begin{abstract}
Classical-quantum computational complexity separations are an important
 motivation for the long-term development of digital quantum computers,
 but classical-quantum complexity equivalences are just as important in
 our present era of noisy intermediate-scale quantum devices
 for framing near-term progress towards quantum supremacy.
We establish one such equivalence using a noisy quantum circuit model
 that can be simulated efficiently on classical computers.
With respect to its noise model, quantum states
 have a robust decomposition into a sequence of operations that each extend the state by one qubit
 without spreading errors between qubits.
This enables universal quantum sampling of states with an efficient representation in this robust form
 and observables with low quantum weight that can be sampled from general measurements
 on a few qubits and computational basis measurements on the remaining qubits.
These robust decompositions are not unique, and we construct two distinct variants,
 both of which are compatible with machine-learning methodology.
They both enable efficiently computable lower bounds on von Neumann entropy
 and thus can be used as finite-temperature variational quantum Monte Carlo methods.
\end{abstract}

\section{Introduction}

Quantum information science (QIS) has recently entered an era of noisy intermediate-scale quantum (NISQ) devices \cite{NISQ}
 that are large enough in size and accurate enough in function to compete favorably against digital classical computers
 at specialized computational tasks designed primarily to favor quantum computers \cite{Google_supremacy}.
The popularity of QIS is increasing in the NISQ era, which is attracting broad interest and activity from people without doctoral training in the subject (myself included).
A lot of this activity focuses on hybrid classical-quantum computations that aspire to improve overall performance
 by offloading key tasks from a digital classical computer to a NISQ device,
 thus delineating a practical boundary between classical and quantum computing.
However, this boundary might be distorted if there is too much emphasis on accommodating NISQ devices
 and not enough emphasis on improving classical algorithms.
As John Preskill cautions about the NISQ era \cite{NISQ}, ``making a quantum circuit more noise resilient may also make it easier to simulate classically''.
It would be doubly counterproductive if this happens and no one realizes it.

From the historical perspective of quantum simulation methodology, NISQ devices are best characterized as a type of variational Monte Carlo (VMC) method \cite{VMC_origin}.
The amount of noise per gate limits the circuit depth of NISQ devices and the relatively slow feedback loop between measurements, classical post-processing, and control
 restrict their operation to tunable quantum state samplers.
As in other VMC methods, the quantum state that they sample is tuned to optimize a function of their observed statistics.
Correspondingly, NISQ device applications favor the quantum approximate optimization algorithm \cite{QAOA} and the variational quantum eigensolver \cite{VQE}
 over more powerful but resource-intensive quantum algorithms such as quantum phase estimation \cite{QPE}.
While there are many realizable NISQ circuits and classical VMC methods and it will be difficult to compare them all precisely,
 the general questions that we ought to ask in fairly assessing their relative utility are:
\begin{enumerate}
\item \label{state_question} What are the differences between their sets of efficiently accessible quantum states?
\item \label{observable_question} What are the differences between their sets of efficiently measurable observables?
\end{enumerate}
Here, efficiency refers to low classical computational costs and high quantum error rates.
We should also be mindful that VMC methods are notoriously expensive for a reason that NISQ devices cannot avoid:
 the formidable number of statistical samples that are required to reduce sampling errors to an acceptable level for practical use.

The purpose of this paper is to present a noisy quantum circuit model that is efficient to simulate on a classical computer
 and a useful reference point for comparisons between NISQ devices and classical VMC methods.
As a hypothetical NISQ device, its gate sets, circuit depths, and noise rates can be compared to physical NISQ devices.
As a classical algorithm, it can be benchmarked against established VMC methods.
These circuits also support a robust mode of operation that can prepare any quantum state accurately,
 but usually with much less efficiency than universal quantum computation (UQC).
This model is intended to complement complexity-theoretic comparisons
 between classical computers and NISQ devices \cite{NISQ_complexity} that
 rely on noise assumptions and complexity conjectures.
Ongoing comparisons such as in Google's recent quantum supremacy experiment \cite{Google_supremacy}
 are using noise to reduce classical computational costs by limiting the number of samples computed
 from otherwise exact classical simulations of noise-free NISQ devices \cite{noise_dilution}.
Although the methods discussed in this paper are not yet mature enough for this application,
 they may eventually serve as a more efficient classical reference for future comparisons with NISQ devices.

\subsection{Noisy quantum complexity}

Much of QIS has been shaped by the detrimental effects of noise and uncertainty on the computational power of quantum systems
 and the protective theoretical frameworks that have been developed to mitigate their effects.
One particularly compelling result of these efforts is the sharp classical-quantum computational boundary
 created by Pauli-stabilizer quantum error correction (QEC) \cite{stabilizer_QEC}.
This form of QEC implements highly reliable logical stabilizer operations -- Clifford circuits combined with state preparation and measurement in the computational basis --
 using noisy physical stabilizer operations and some classical computation to decode errors.
If the noise can be represented as a statistical combination of stabilizer operations,
 then this entire process can be simulated efficiently on a classical computer using stabilizer simulations \cite{stabilizer_simulation}.
UQC is enabled by introducing noisy physical non-Clifford gates such as the $T$ gate with depolarizing noise \cite{magic_state_distillation}.
In this case, there is a sharp noise threshold, above which the noise can be statistically unravelled as stabilizer operations and efficiently simulated on a classical computer
 and below which the noise can be distilled to form highly reliable logical $T$ gates that enable UQC.
Unfortunately, this sharp classical-quantum boundary is not relevant to the NISQ era because the QEC noise thresholds
 are too low and the number of physical qubits per logical qubit is too high.

Efforts to establish a classical-quantum computational boundary in the NISQ era have focused on sampling problems.
The distribution of computational basis measurements for quantum states that can be prepared efficiently on NISQ devices
 sometimes correspond to probability functions that cannot be evaluated efficiently on classical computers.
Popular proposals include random quantum circuits \cite{random_sampling}, instantaneous quantum polynomial (IQP) circuits \cite{IQP_sampling}, and photon-resolved linear-optical networks \cite{boson_sampling}.
However, verification has a high sampling complexity in general \cite{supremacy_verification},
 and specific proposals may exceed the available experimental capabilities \cite{boson_noise}
 or need redundancy and classical error correction \cite{IQP_noise}.
Even in a successful experimental realization \cite{Google_supremacy}, its overlap with the ideal output distribution decays exponentially in the circuit depth and the number of qubits,
 although this decay is slow enough to be outpaced by an exponentially growing cost of its simulation on classical computers.
The theoretical frameworks underlying these efforts are focused on asymptotic computational complexity separations
 and missing a counterpart to the noisy Clifford+$T$ circuit model that characterizes a transition, either sharp or gradual, from a noisy regime
 with a classical-quantum computational equivalence to a low-noise regime where quantum computing devices
 have a clear computational advantage over classical computers.

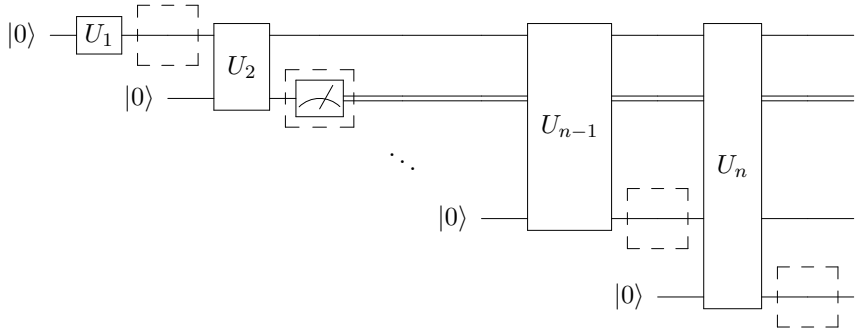
\begin{figure}
\centering
$\scalebox{0.9}{\Qcircuit @C=1em @R=1em {
\lstick{\ket{0}} & \gate{U_1} &        \qw & \multigate{1}{U_2} & \qw      & \qw     & \qw                & \multigate{3}{U_{n-1}} & \qw                  & \multigate{4}{U_n} & \qw & \qw \\
                       &                   & \lstick{\ket{0}} & \ghost{U_2}           & \meter & \cw     & \cw                & \pureghost{U_{n-1}} \cw    & \cw                  & \pureghost{U_n}     \cw & \cw & \cw \\
                       &                   & \push{\rule{1.5em}{0em}} &               &            & \ddots &                       & \pureghost{U_{n-1}}      & \push{\rule{1.5em}{0em}} & \pureghost{U_n}        & \push{\rule{1.5em}{0em}}  & \\
 \push{\rule{0em}{2em}} &   &                        &                              &        & \push{\rule{2.5em}{0em}} & \lstick{\ket{0}} & \ghost{U_{n-1}}           & \qw                 & \ghost{U_n}           & \qw & \qw \\
 \push{\rule{0em}{2em}}  &  &                        &                              &            &            &  \push{\rule{1.5em}{0em}} &  & \lstick{\ket{0}} & \ghost{U_n}           & \qw & \qw
\gategroup{1}{3}{1}{3}{2.3em}{--}
\gategroup{2}{5}{2}{5}{0.8em}{--}
\gategroup{4}{9}{4}{9}{2.3em}{--}
\gategroup{5}{11}{5}{11}{2.3em}{--}
}}$
\caption{\label{noisy_circuit_diagram} An $n$-qubit example of our noisy quantum circuit model.
Dashed boxes denote the possible locations where errors can occur, and an error is depicted on the second qubit. }
\end{figure}

In this paper, we consider a noisy quantum circuit model that uses noise to control its classical simulation costs
 while providing more flexibility than Clifford+$T$ circuits for the typical sampling tasks of NISQ devices.
As shown in Fig.\@ \ref{noisy_circuit_diagram}, the model alternates between initializing a new qubit,
 performing a quantum operation on all available qubits,
 and then applying a noise channel to the new qubit that either does nothing and returns the qubit
 or measures it in the computational basis and returns the measurement outcome instead of the qubit.
A probability distribution over qubit clusters that survive to the end of the circuit defines its noise statistics.
As the average size of the surviving qubit cluster grows,
 it becomes more efficient to sample from an idealized NISQ device than to simulate it on a classical computer.
Compared with physical noise in a realistic NISQ device, this noise
 is both less onerous because errors are heralded and very restricted in location
 and more onerous because errors permanently remove qubits for the remainder of the circuit.
In this model, the classical simulation cost of quantum systems is controlled by restricting their size rather than their complexity.
There are other ways to control cost such as restricting the
 bond dimension in a matrix-product-state representation \cite{bond_dimension}
 or amount of negativity in a quasiprobability representation \cite{quasiprobability},
 but these quantities are more difficult to control with noisy quantum operations
 and rely on more specialized complexity assumptions that
 apply to specific quantum state representations rather than generic quantum states.

\subsection{Quantum robustness}

The term ``robust'' appears often in the QIS literature with varying specificity and
 usually refers to an insensitivity of one quantity to variations in another quantity.
The frequency of its use signifies its compelling role in the narrative of quantum information's struggle against noise and uncertainty.
At an abstract level, precise definitions for the robustness of states \cite{robust_state} and gates \cite{robust_gate} have been suggested
 to quantify their ability to preserve and generate bipartite entanglement with respect to specific noise models.
At a physical level, robust gates of specific quantum computing devices have been designed to minimize their
 sensitivity to the largest known sources of uncertainty \cite{robust_operation},
 and quantum control theory \cite{robust_control} has adopted concepts of robust control
 from the broader control theory literature \cite{robust_control_book}.
At an algorithmic level, robust quantum algorithms have been developed to maintain an
 advantage over classical algorithms in the presence of tailored artificial noise models,
 such as evaluating Boolean functions with noisy inputs \cite{robust_polynomial} 
 and learning parity from an oracle with noisy outputs \cite{robust_learning}.
In some cases, robustness can simply refer to the limitations of a quantum computing device being proportionate
 to limited expectations of its utility, such as with analog quantum simulation of noisy and uncertain physical systems \cite{robust_simulation}.

\begin{figure}
\centering
(a)
$\scalebox{0.9}{\Qcircuit @C=0.5em @R=0.5em {
& \multiprepareC{5}{\ket{\psi_{n+1}}} & \qw & & \multiprepareC{4}{\ket{\psi_{n}}} & \qw & \ctrl{5} & \qw & \qw & \ctrl{5} & \qw & \ctrl{5} & \qw \\
& \pureghost{\ket{\psi_{n+1}}} & \qw & & \pureghost{\ket{\psi_{n}}} & \qw & \qw & \ctrl{4} & \qw & \qw & \ctrl{4} & \ctrl{4} & \qw \\
& \pureghost{\ket{\psi_{n+1}}} & \push{\raisebox{0.5em}{\vdots} \rule{0.8em}{0em}} & \! \! \! \raisebox{-2.1em}{=} & \pureghost{\ket{\psi_{n}}} & \raisebox{0.5em}{\vdots} & & & \push{\cdots} & & & & \push{\rule{0.7em}{0em}} \\
& \pureghost{\ket{\psi_{n+1}}} & \qw & \push{\rule{1.5em}{0em}} & \pureghost{\ket{\psi_{n}}} & \qw & \qw & \qw & \qw & \ctrl{2} & \ctrl{2} & \ctrl{2} & \qw \\
& \pureghost{\ket{\psi_{n+1}}} & \qw & & \pureghost{\ket{\psi_{n}}} & \qw & \qw & \qw & \qw & \ctrl{1} & \ctrl{1} & \ctrl{1} & \qw  \\
& \pureghost{\ket{\psi_{n+1}}} & \qw & & \lstick{\ket{0}} & \gate{U_0^{(n+1)}} & \gate{U_1^{(n+1)}} & \gate{U_2^{(n+1)}} & \qw & \gate{U_{2^n - 3}^{(n+1)}} & \gate{U_{2^n - 2}^{(n+1)}} & \gate{U_{2^n - 1}^{(n+1)}} & \qw
}}$\\
\vspace{1.5em}(b)
$\scalebox{0.9}{\Qcircuit @C=0.5em @R=0.5em {
& \multiprepareC{5}{\ket{\psi_{n+1}}} & \meter & \cw & & \multiprepareC{4}{\ket{\psi_{n}}} & \meter & \control \cw & \cw & \cw & \control \cw & \cw & \control \cw & \cw & \cw \\
& \pureghost{\ket{\psi_{n+1}}} & \meter & \cw & & \pureghost{\ket{\psi_{n}}} & \meter & \cwx \cw & \control \cw & \cw & \cw \cwx & \control \cw & \control \cwx \cw & \cw & \cw \\
& \pureghost{\ket{\psi_{n+1}}} & \raisebox{0.5em}{\vdots} & & \raisebox{-2.5em}{=} & \pureghost{\ket{\psi_{n}}} & \raisebox{0.5em}{\vdots} & \cwx & \cwx & \push{\cdots}  & \cwx & \cwx & \cwx &  \\
& \pureghost{\ket{\psi_{n+1}}} & \meter & \cw & \push{\rule{1.5em}{0em}} & \pureghost{\ket{\psi_{n}}} & \meter & \cwx \cw & \cwx \cw & \cw & \control \cwx \cw & \control \cwx \cw & \control \cwx \cw & \cw & \cw \\
& \pureghost{\ket{\psi_{n+1}}} & \meter & \cw & & \pureghost{\ket{\psi_{n}}} & \meter & \cwx \cw & \cwx \cw & \cw & \control \cwx \cw & \control \cwx \cw & \control \cwx \cw & \cw & \cw \\
& \pureghost{\ket{\psi_{n+1}}} & \meter & \cw & & \lstick{\ket{0}} & \gate{U_0^{(n+1)}} & \gate{U_1^{(n+1)}} \cwx & \gate{U_2^{(n+1)}} \cwx & \qw & \gate{U_{2^n - 3}^{(n+1)}} \cwx & \gate{U_{2^n - 2}^{(n+1)}} \cwx & \gate{U_{2^n - 1}^{(n+1)}} \cwx & \meter & \cw 
}}$
\caption{\label{robust_UQS_circuits} (a) State preparation and (b) computational basis measurement for a general quantum state decomposed into a recursive circuit \cite{conditional_quantum}.
For the noisy circuit model shown in Fig.\@ \ref{noisy_circuit_diagram}, this corresponds to (a) no errors and (b) an error on every qubit. }
\end{figure}
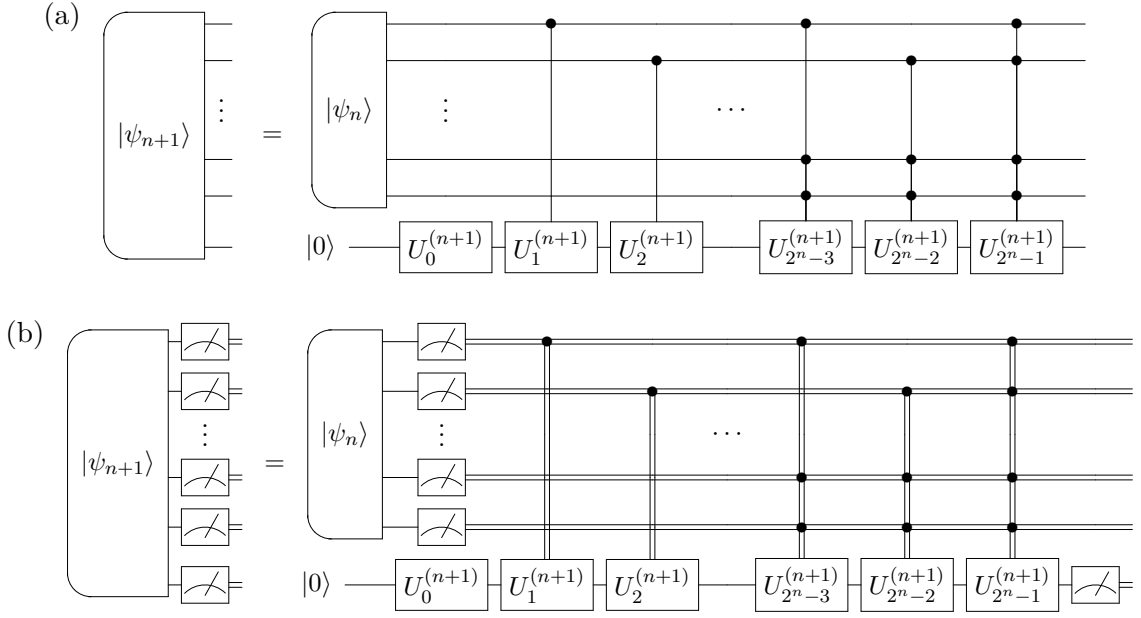

For the purposes of this paper, we consider a specific strong notion of ``robust'' that is suitable for a heralded subsystem error model
 -- robust operations are those that do not spread errors beyond the subsystems in which they occur.
This is a stronger requirement than fault-tolerance criteria in QEC \cite{FTQEC} that do not constrain the location or spreading of physical errors
 as long as logical errors can be corrected, and it substantially limits the set of robust operations.
However, there are useful examples of robust operations such as
 the quantum Bayesian networks \cite{quantum_bayesian} shown in Fig.\@ \ref{robust_UQS_circuits}.
It is straightforward to decompose any quantum state into such circuits \cite{conditional_quantum},
 and we refer to their implementation within the noisy quantum circuit model as \textit{universal quantum sampling} (UQS).
UQC can implement any unitary operation, some more efficiently than others, and select the most efficient one to prepare a target state,
 whereas UQS can implement only a very limited set of unitary operations that can still prepare any state, but sometimes much less efficiently than UQC.
The generic UQS circuit in Fig.\@ \ref{robust_UQS_circuits} has a depth that is exponential in its number of qubits,
 and only special instances that can either remove or somehow compress many of its gates will be useful in practice.
Unlike QEC, this robust UQS implementation has no space or time overhead, and it is not restricted to a discrete set of elementary gates.
Its underlying structure of replacing quantum controlled gates with measurements and classical controls
 is a general characteristic of semiclassical algorithms in quantum computation \cite{semiclassical_algorithms}.

Even before analyzing it in more detail, robust UQS helps to answer questions \ref{state_question} and \ref{observable_question}
 by establishing a wide region of equivalence between classical computers and an idealized NISQ device defined by our noisy quantum circuit model.
Its efficiently accessible states are those in which Fig.\@ \ref{robust_UQS_circuits} is sparse or structured,
 and its efficiently measurable observables are those that can be measured on qubit clusters with a high survival probability.
We will refer to the number of qubits that need to be measured outside of the computational basis
 in order to measure an observable as its \textit{quantum weight}.
Since smaller clusters typically have higher survival probabilities,
 it is the observables with low quantum weight that are efficiently measurable.
While we seek to expand and clarify this region of equivalence, we
 cannot discount the existence of other regions of equivalence or other possibilities beyond it.
Classical algorithms and realistic NISQ devices may have efficient access to states and observables
 that are not efficiently accessible to robust UQS.

\subsection{VMC methods}

When narrowly defined, VMC methods refer to a specific set of mathematical ingredients
 but a wide variety of variational forms ranging from simple products of pairwise functions \cite{VMC_origin}
 to sophisticated machine learning models \cite{MLMC}.
VMC needs an unnormalized state $|\psi\rangle$, a target observable $B$, and an orthonormal basis $|i\rangle$
 in which $\langle i |\psi\rangle$ and $\langle i | B | j \rangle$ are efficient to compute and $\langle i | B | j \rangle$ is sparse.
The expectation value of $B$ is sampled by VMC as
\begin{equation} \label{VMC_sampling}
 \frac{\langle \psi | B | \psi \rangle}{\langle \psi | \psi \rangle} = \sum_{i} b(i) p(i) \ \ \ \mathrm{for} \ \ \ b(i) = \sum_j \langle i | B | j \rangle \frac{\langle j | \psi \rangle}{\langle i | \psi \rangle} \ \ \ \mathrm{and} \ \ \ p(i) = \frac{| \langle i | \psi \rangle |^2}{\langle \psi | \psi \rangle} ,
\end{equation}
 where an efficiently computable estimator $b(i)$ is sampled according to a probability $p(i)$
 with Markov-chain Monte Carlo (MCMC) methods such as the Metropolis algorithm \cite{MCMC}.
The essential feature of MCMC sampling is that only $p(i)/p(j)$ is required to be efficiently computable rather than $p(i)$.
The variational aspect occurs when $B$ is a Hamiltonian and parameters in $|\psi\rangle$ are tuned to minimize its expectation value.
Free energy minimization requires the additional computation of von Neumann entropy,
 which in most cases cannot be described as the expectation value of an efficiently computable $B$.
A variational upper bound on free energy only needs lower bounds on von Neumann entropy,
 but even bounds are difficult to compute in an efficient and systematically improvable manner.
Thus, there are not yet any broadly applicable finite-temperature VMC methods.

More broadly defined, VMC may include other quantum Monte Carlo (QMC) methods that use the variational principle
 to adjust their compromise between bias and variance in mitigating ``sign problems'' \cite{sign_problem}.
Abstractly, QMC methods arrange expectation values as a ratio of series with efficiently computable terms
 and adapt them for MCMC sampling as
\begin{equation}
 \frac{\sum_i f(i)}{\sum_i g(i)} = \frac{\sum_i b(i) p(i)}{\sum_i \mathrm{sgn}(g(i)) p(i)} \ \ \ \mathrm{for} \ \ \ b(i) = \frac{f(i)}{|g(i)|} \ \ \ \mathrm{and} \ \ \ p(i) = \frac{|g(i)|}{\sum_j | g(j) |} .
\end{equation}
When $g(i)$ is non-negative, this calculation is comparable to Eq.\@ (\ref{VMC_sampling}).
Otherwise, sampling from the denominator can have high variance caused by cancellations between $g(i)$ values of opposite sign.
In particular, its variance in fermionic systems often grows exponentially with the number of fermions.
Such variance problems are a consequence of constructing unbiased sampling methods,
 typically using imaginary-time evolution at zero temperature \cite{DMC}
 and path integrals at finite temperature \cite{PIMC}.
At zero temperature, there are methods such as the fixed-node approximation \cite{constrained_DMC} 
 to mitigate the growth of variance and bias the statistics while still sampling from a valid state that simply differs from the ground state.
Thus an upper bound on the ground state energy can be maintained, and the variational minimization of energy is a guide for optimizing the constrained nodal structure.
Similar fixed-node approximations can be used at finite temperature \cite{contrained_PIMC},
 but again free energy is not efficiently accessible as a variational guide for systematically reducing bias.

The important special case of VMC that we explore in this paper is when calculations of $|\psi\rangle$ in Eq.\@ (\ref{VMC_sampling})
 correspond to the direct simulation of physical quantum processes.
As in the variational quantum eigensolver \cite{VQE}, we would sample the expectation value of $B$ in a basis $|i\rangle$ that diagonalizes it
 and directly generate independent samples from $p(i)$ instead of having to wait for the mixing time of an MCMC method.
The underlying variance in Eq.\@ (\ref{VMC_sampling}) will not change,
 but the rate at which independent samples are generated can be dramatically accelerated.
Such behavior was observed in a recent VMC method based on autoregressive neural networks \cite{autoregressive_VMC},
 which fortuitously corresponds to an efficient instance of Fig.\@ \ref{robust_UQS_circuits}.
Whereas much of QMC method development has been moving away from VMC and exploring diverse strategies for minimizing bias while mitigating sign problems,
 there are substantial benefits to imposing further restrictions on VMC instead.
In particular, it becomes possible to design efficiently computable lower bounds on von Neumann entropy
 using quantum Shannon theory \cite{quantum_Shannon_theory} when a variational state is the outcome of a physical quantum process.
There are already some isolated examples of such entropy bounds \cite{entropy_bound} in addition to 
 exact calculations of Shannon entropy from variational classical many-body states \cite{autoregressive_classical}.
Thus we establish a technical path to finite-temperature VMC methods.

\section{Robust decompositions}\label{decomposition_section}

We use a recursive strategy and notation for the construction and analysis of robust state decompositions.
A Hilbert space of composite dimension $N = N_A N_B$ is partitioned into a pair of subsystems, labelled $A$ and $B$, of dimensions $N_A$ and $N_B$ respectively.
We utilize tensor-product notation to write states and operations on this Hilbert space.
For example, a general state $\rho$ is expanded into matrix elements $\rho_{ik,jl} \coloneqq ( \langle i | \otimes \langle k| ) \rho ( | j \rangle \otimes | l \rangle )$ as
\begin{equation} \label{general_rho}
 \rho = \sum_{i,j = 0}^{N_A - 1} \sum_{k,l = 0}^{N_B - 1} \rho_{ik,jl} |i\rangle \langle j| \otimes |k\rangle \langle l | .
\end{equation}
In each case, we decompose $\rho$ into a state $\sigma$ on subsystem $B$
 and a robust reconstruction operation $R$ that extends states from subsystem $B$ to the full Hilbert space,
\begin{equation} \label{general_sigma}
 \rho = R(\sigma) \ \ \ \mathrm{for} \ \ \ \sigma = \sum_{i,j=0}^{N_B - 1} \sigma_{i,j} |i\rangle \langle j|,
\end{equation}
 for various restrictions on $R$, $\rho$, and $\sigma$.
The complete decomposition is recursively derived by partitioning subsystem $B$ into two subsystems and decomposing $\sigma$ the same way as $\rho$.
Each $A$ will be an elementary subsystem such as one qubit in the noisy circuit model,
 and each $B$ will be a shrinking set of already-prepared elementary subsystems.

\subsection{Conditional probability decomposition}\label{probability_decomposition_subsection}

When $\rho$ is a classical state that is diagonal in the computational basis,
 we are able to use the standard concept of conditional probabilities as a robust decomposition.
A classical $\rho$ is decomposed into a classical $\sigma$ and $R$ as
\begin{align} \label{classical_decomposition}
 \rho &= \sum_{i= 0}^{N_A-1} \sum_{j = 0}^{N_B-1} P_{AB}(i,j) |i\rangle \langle i| \otimes |j\rangle \langle j | = R(\sigma) \ \ \ \mathrm{for} \notag \\
  R(\rho_B) &= \sum_{i= 0}^{N_A-1} \sum_{j = 0}^{N_B-1} P_{A|B}(i|j) |i\rangle \langle i| \otimes |j\rangle \langle j | \rho_B |j\rangle \langle j | 
  \ \ \ \mathrm{and} \ \ \ \sigma = \sum_{i = 0}^{N_B-1} P_B(i) |i\rangle \langle i| ,
\end{align}
 which respectively correspond to marginal and conditional probabilities,
\begin{equation}
 P_B(i) = \sum_{j=0}^{N_A-1} P_{AB}(j,i) \ \ \ \mathrm{and} \ \ \ P_{A|B}(i|j) = \frac{P_{AB}(i,j)}{P_B(j)}.
\end{equation}
While we would usually interpret $P_{A|B}(i|j)$ as a set of conditional states parameterized by $j$,
 its physical interpretation here is as an operation rather than a set of states.
Similarly, we can also interpret marginalization as an operation $M$ from the full Hilbert space
 back to subsystem $B$, $\sigma = M(\rho)$, defined as
\begin{equation}
 M(\rho_A \otimes \rho_B) = \sum_{i= 0}^{N_A-1} \sum_{j = 0}^{N_B-1} \langle i| \rho_A | i\rangle \otimes |j\rangle \langle j | \rho_B |j\rangle \langle j |.
\end{equation}
While $R$ depends on the choice of $\rho$, $M$ is the same for every decomposed state.

An important property of the conditional probability decomposition is the chain rule of entropy, $S(\rho) \coloneqq - \mathrm{tr} ( \rho \ln \rho)$,
 which decomposes the entropy of $\rho$ into the entropies of $\sigma$ and an average entropy of $R$ applied to computation basis states,
\begin{equation}
 S(\rho) = S(\sigma) + \sum_{i=0}^{N_B-1} P_B(i) S(R(|i\rangle \langle i|)) . \label{entropy_chain_rule}
\end{equation}
Thus we can efficiently sample the entropy of $\rho$ by sampling $S(R(|i\rangle \langle i|))$
 and similar terms corresponding to all of the reconstruction operations used to construct $\sigma$ recursively.
The evaluation of $S(R(|i\rangle \langle i|))$ and sampling from $P_B(i)$ are efficient if all $P_{A|B}(i | j)$ throughout
 the recursive decomposition can be efficiently evaluated for any $j$ to facilitate conditional sampling over $i$.
Continuing our operational interpretation of conditional probabilities,
 we can interpret the second term on the right-hand side of Eq.\@ (\ref{entropy_chain_rule})
 as the entropy produced by $R$ acting on $\sigma$ in addition to being the entropy of subsystem $A$ conditioned on $B$.

The decomposition in Eq.\@ (\ref{classical_decomposition}) is unique for a given $P_{AB}$,
 but other decompositions are possible.
Our robustness criterion does not constrain classical operations performed in the computational basis
 because they are not affected by our error model.
Furthermore, if we consider measurement-collapsed qubits as adjustable bits rather than static measurement records,
 then a more general $R$ could change the marginal state on subsystem $B$.
We can interpret such an $R$ as a non-trivial classical operation on subsystem $B$ conditioned on the value assigned to subsystem $A$.
With this more general form for $R$, the decomposition in Eq.\@ (\ref{classical_decomposition})
 is no longer unique, and we lose direct access to an efficiently computable product structure, $P_{AB} = P_{A|B} P_B$,
 from which to calculate entropy efficiently using Eq.\@ (\ref{entropy_chain_rule}).
This increased difficulty in calculating entropy is one consequence of $R$ losing its reversibility,
 which we will explore further in Sec.\@ \ref{density_decomposition_subsection}.
Therefore we have both conceptual and practical reasons for restricting classical state decompositions to the form in Eq.\@ (\ref{classical_decomposition}).

For recursive state decompositions to be computationally useful, we need an efficiently computable model for each $P_{A|B}$.
Historically, Bayesian networks \cite{Bayesian_networks} have been the most common models, whereby the conditional dependencies of subsystem $A$
 are just restricted to a small subsystem contained within subsystem $B$.
However, Bayesian networks are not able to represent ``diffuse'' conditionings in which one subsystem depends on all previous subsystems,
 albeit in a simple manner.
It is now popular to apply more flexible machine-learning models to correct such deficiencies \cite{autoregressive_classical}.
These models are usually unconstrained real-valued functions that need to be constrained for specific applications.
For example,
\begin{equation}
 P_{A|B}(i | j) = \frac{\exp (f(i,j)) }{\sum_{k=0}^{N_A-1} \exp ( f(k,j) ) }
\end{equation}
 maps an unconstrained function $f$ to a non-negative and properly normalized $P_{A|B}$ with
 a form reminiscent of a Boltzmann distribution.
A truncated cluster expansion of $f$ guided by the substructure of subsystem $B$
 might be a useful model of intermediate complexity between Bayesian networks and machine-learning models
 with a form reminiscent of the typical few-body structure of physical many-body Hamiltonians.

\subsection{Conditional quantum decomposition}\label{quantum_decomposition_subsection}

When $\rho$ is a pure quantum state,
\begin{equation}
 \rho = |\psi_{AB}\rangle \langle \psi_{AB}| \ \ \ \mathrm{for} \ \ \ |\psi_{AB}\rangle = \sum_{i=0}^{N_A-1} \sum_{j=0}^{N_B-1} \psi_{AB}(i,j) |i\rangle \otimes |j\rangle,
\end{equation}
 we can construct a robust decomposition as an isometric extension of another pure state,
\begin{equation} \label{pure_decomposition}
 \rho = R(\sigma) = V |\psi_B \rangle \langle \psi_B | V^\dag ,
\end{equation}
 where $\sigma$ is constructed from a non-standard marginalization with an arbitrary phase $\theta$,
\begin{equation} \label{pure_sigma}
 \sigma = |\psi_B\rangle \langle \psi_B | \ \ \ \mathrm{for} \ \ \ |\psi_B \rangle = \sum_{i=0}^{N_B-1} \psi_B(i) |i\rangle \ \ \ \mathrm{and} \ \ \ \psi_B(j) = e^{i\theta(j)}\sqrt{ \sum_{k=0}^{N_A-1} |\psi_{AB}(k,j) |^2 },
\end{equation}
 and $R$ contains a quantum analog of conditional probability \cite{quantum_bayesian},
\begin{equation} \label{pure_R}
 R(\rho_B) = V \rho_B V^\dag \ \ \ \mathrm{for} \ \ \ V = \sum_{i=0}^{N_A-1} \sum_{j=0}^{N_B-1} \psi_{A|B}(i | j ) | i \rangle \otimes |j\rangle \langle j| \ \ \ \mathrm{and} \ \ \ \psi_{A|B}(i | j ) = \frac{\psi_{AB}(i,j)}{\psi_B(j)}.
\end{equation}
The isometry $V$ alternatively can be split into the preparation of subsystem $A$ in a pure state followed by a unitary acting on subsystem $A$ controlled by subsystem $B$.
Whereas $R$ in Eq.\@ (\ref{classical_decomposition}) is reversible on classical states, $R$ in Eq.\@ (\ref{pure_R}) is now also reversible on quantum states.
Its ``marginalization'' operation $M$ now depends on $R$ as $M = R^\dag$.

The functions that might be interpreted as marginal and conditional quantum states, $\psi_B$ and $\psi_{A|B}$,
 are not unique because of the arbitrary phase $\theta$ in Eq.\@ (\ref{pure_sigma}).
The conditional quantum state can be physically interpreted as a pure state in subsystem $A$ conditioned on a computational basis measurement outcome in subsystem $B$.
In this interpretation, $\theta$ is just an arbitrary global phase assigned to each conditional state.
Correspondingly, the only physically meaningful property of $|\psi_B\rangle$
 is the distribution of its computational basis measurement outcomes, which does not depend on $\theta$.
We can make the decomposition in Eq.\@ (\ref{pure_decomposition}) unique by prescribing an arbitrary unique phase such as $\theta = 0$.

Just as in the classical case, a variety of models can be used to represent $\psi_{A|B}$,
 from Bayesian networks \cite{quantum_bayesian} to machine-learning models \cite{autoregressive_VMC}.
Again, we can apply the correct physical constraints to an unconstrained function $f$ with various mappings such as
\begin{equation}
 \psi_{A|B}(i | j) = \frac{f(i,j)}{\sqrt{\sum_{k=0}^{N_A-1} | f(k,j) |^2}},
\end{equation}
 where $f$ is now complex-valued in the quantum case.
The constraint of $\theta = 0$ in Eq.\@ (\ref{pure_sigma})
 corresponds to a real-valued $f$ in each prior $\psi_{A|B}$ of a recursive decomposition,
 effectively deferring the introduction of non-trivial quantum phases until a state is extended onto its final subsystem.
In Fig.\@ \ref{robust_UQS_circuits}, this would correspond to all $U_n^{(m)}$ gates being $X$ rotations and
 appending arbitrary controlled-$Z$ rotations to the final state preparation circuit.
A more balanced variational form should probably add phases with each step of the decomposition
 to build up all relevant correlations with new subsystems as they are introduced.

\subsection{Conditional density decomposition} \label{density_decomposition_subsection}

By inspecting Eqs.\@ (\ref{classical_decomposition}) and (\ref{pure_R}), we can infer a more general form of robust operation,
\begin{equation} \label{general_R}
 R(\rho_B) = \sum_{i,j = 0}^{N_A-1} \sum_{k,l = 0}^{N_B-1} R_{ik,jl} |i\rangle \langle j| \otimes |k\rangle \langle k | \rho_B |l\rangle \langle l | ,
\end{equation}
 for which the classical and pure-quantum states are special cases. 
This is the most general form of $R$ that preserves computational basis measurements in subsystem $B$.
Otherwise, the form of $R$ could be generalized further to include any terms that do not couple between diagonal and off-diagonal matrix elements of the input state.
There are no obvious benefits to this more general form, and we do not consider it any further in this paper.
We require that the $R$ in Eq.\@ (\ref{general_R}) be completely positive and trace preserving, which corresponds to
 $R_{ij,kl}$ being a Hermitian positive semidefinite matrix with $ij$ as its row index and $kl$ as its column index
 while also satisfying the normalization condition
\begin{equation} \label{trace_preservation}
 \sum_{i=0}^{N_A-1} R_{ij,ij} = 1 \ \ \ \mathrm{for} \ \ \ 0 \le j \le N_B - 1.
\end{equation}
These conditions are similar to the semidefinite and trace constraints on $\rho_{ij,kl}$ and $\sigma_{i,j}$.

We can prove the existence of robust decompositions in the general quantum case, but they are again highly non-unique.
From the general forms of $\rho$, $\sigma$, and $R$ in Eqs.\@ (\ref{general_rho}), (\ref{general_sigma}), and (\ref{general_R}),
 we can trivially solve for $R$ as a function of $\rho$ and $\sigma$,
\begin{equation} \label{solve_for_R}
 \rho = R( \sigma ) \ \ \ \mathrm{for} \ \ \ R_{ij,kl} = \frac{\rho_{ij,kl}}{\sigma_{j,l}}.
\end{equation}
The trace preservation condition in Eq.\@ (\ref{trace_preservation}) uniquely defines the diagonals of $\sigma$ to be
\begin{equation}
 \sigma_{i,i} = \sum_{j=0}^{N_A-1} \rho_{ji,ji} .
\end{equation}
We can then guarantee that $R_{ij,kl}$ is Hermitian positive semidefinite with the choice
\begin{equation} \label{rank1_sigma}
 \sigma_{i,j} = \sqrt{ \sum_{k=0}^{N_A-1} \rho_{ki,ki} } \sqrt{ \sum_{k=0}^{N_A-1} \rho_{kj,kj} }
\end{equation}
because $R_{ij,kl}$ in Eq.\@ (\ref{solve_for_R}) is then a symmetric diagonal rescaling of $\rho_{ij,kl}$,
 which preserves the Hermitian positive semidefinite structure of $\rho_{ij,kl}$ itself.
In the typical case, all of the eigenvalues of $\rho_{ij,kl}$ and $R_{ij,kl}$ will be nonzero, and any perturbation of $\sigma_{i,j}$ from Eq.\@ (\ref{rank1_sigma})
 that remains inside the convex set preserving its diagonal elements and Hermitian positive
 semidefinite structure corresponds to a valid perturbation of $R_{ij,kl}$ and the overall robust decomposition.
When the altered $\sigma_{i,j}$ also has only nonzero eigenvalues,
 it has a manifold of perturbations with dimension $N_B(N_B-1)/2$ that preserve its Hermitian structure and diagonal elements
 and thus locally define a manifold of valid robust decompositions.

When $\rho$ is no longer a pure state, it becomes desirable to calculate lower bounds on its von Neumann entropy, $S(\rho)$.
We construct lower bounds using quantum relative entropy, $S(\sigma \| \sigma_0) \coloneqq \mathrm{tr}(\sigma \ln \sigma - \sigma \ln \sigma_0)$,
 and its monotonicity under any quantum operation $R$ \cite{quantum_Shannon_theory},
\begin{equation} \label{monotonicity}
 S(\sigma \| \sigma_0) \ge S(R(\sigma) \| R(\sigma_0)).
\end{equation}
The basic strategy is to use $R$ from Eq.\@ (\ref{general_R}) and choose $\sigma_0$ to enable efficient computation of $\ln \sigma_0$.
For example, if $\sigma_0$ is the dephased form of $\sigma$,
 we can derive an analog of Eq.\@ (\ref{entropy_chain_rule}),
\begin{equation} \label{quantum_entropy_bound1}
 S(\rho) \ge S(\sigma) + \sum_{i=0}^{N_B-1} \langle i | \sigma | i \rangle S(R(|i\rangle \langle i|)) .
\end{equation}
The inequality in Eq.\@ (\ref{monotonicity}) is well understood within quantum Shannon theory \cite{quantum_Shannon_theory},
 and it can be saturated only when a common quantum operation is able to reverse the effects of $R$ acting on both $\sigma$ and $\sigma_0$.
If there are saturating operations, one can be written as
\begin{equation}
 M(\rho_{AB}) = \sigma_0^{1/2} R^\dag( R(\sigma_0)^{-1/2} \rho_{AB} R(\sigma_0)^{-1/2} ) \sigma_0^{1/2},
\end{equation}
 which is the Petz recovery map for $R$ acting on $\sigma_0$ that guarantees $M(R(\sigma_0)) = \sigma_0$.
If we interpret the composed operation, $E \coloneqq M \circ R$, as an error channel and apply Eq.\@ (\ref{monotonicity}) to $M$,
 then we can view the looseness of the bound as quantifying the distortion of $\sigma$ by $E$,
\begin{equation} \label{sandwich_bounds}
 S(\sigma \| \sigma_0) \ge S(R(\sigma) \| R(\sigma_0)) \ge S(E(\sigma) \| \sigma_0).
\end{equation}
Thus we must carefully choose $\sigma_0$ to balance the cost and tightness of entropy bounds.
A similar approach can bound the entropy of Markovian matrix-product states \cite{entropy_bound}.

Unlike the classical and pure-quantum cases, $R$ in Eq.\@ (\ref{general_R}) is not generally reversible,
 and we systematically tighten the entropy bounds in Eq.\@ (\ref{sandwich_bounds}) by
 increasing the similarity between $\sigma_0$ and $\sigma$ rather than making $R$ more reversible.
From the limited outputs of our noisy circuit model, an accessible choice for $\sigma_0$ is $\sigma$ conditioned on
 a partial computational basis measurement.
We partition $B$ into a large subsystem $B'$ and a small subsystem $\overline{B'}$
 and consider $\sigma_0$ resulting from computational basis measurements on $B'$,
\begin{equation} \label{repartition}
 \sigma = \sum_{i,j=0}^{N_{B'}-1} \overline{\sigma}_{i,j} \otimes | i \rangle \langle j | \ \ \ \mathrm{and} \ \ \
  \sigma_0 = \sum_{i=0}^{N_{B'}-1} \overline{\sigma}_{i,i} \otimes | i \rangle \langle i | ,
\end{equation}
 where $N_{B'}$ is the dimension of subsystem $B'$ and $\overline{\sigma}_{i,j}$ are the operator components of $\sigma$ on subsystem $\overline{B'}$.
We can similarly define $A' \coloneqq A \cup \overline{B'}$ and repartition $R$ in Eq.\@ (\ref{general_R}) as
\begin{equation} \label{R_repartition}
 R( \rho_{\overline{B'}} \otimes \rho_{B'} ) = \sum_{i,j=0}^{N_{B'}-1} \overline{R}_{i,j}(\rho_{\overline{B'}}) \otimes | i \rangle \langle i | \rho_{B'} | j \rangle \langle j | ,
\end{equation}
 where $\overline{R}_{i,j}$ are superoperators on subsystem $\overline{B'}$ and valid quantum operations when $i=j$.
In this notation, we can describe the effect of the error channel $E$ on $\sigma$ as
\begin{align} \label{error_channel}
 E(\sigma) &= \sum_{i,j=0}^{N_{B'}-1} \widetilde{\sigma}_{i,j} \otimes  | i \rangle \langle j | \ \ \ \mathrm{for} \notag \\
 \widetilde{\sigma}_{i,j} &= \overline{\sigma}_{i,i}^{1/2} \overline{R}_{j,i}^\dag ( \overline{R}_{i,i}(\overline{\sigma}_{i,i})^{-1/2} \overline{R}_{i,j}(\overline{\sigma}_{i,j})
 \overline{R}_{j,j}(\overline{\sigma}_{j,j})^{-1/2} ) \overline{\sigma}_{j,j}^{1/2}.
\end{align}
Notably, $\widetilde{\sigma}_{i,i} = \overline{\sigma}_{i,i}$ because $\overline{R}_{i,i}^\dag (I) = I$ for trace-preserving operations.
Thus the diagonal components of $\sigma$ on subsystem $\overline{B'}$ are preserved by $E$,
 and the overall distortion of $\sigma$ can be reduced by expanding $\overline{B'}$.
The tighter version of Eq.\@ (\ref{quantum_entropy_bound1}) using $\sigma_0$ from Eq.\@ (\ref{repartition}) is
\begin{equation}\label{quantum_entropy_bound2}
 S(\rho) \ge S(\sigma) + \sum_{i=0}^{N_{B'}-1}  \mathrm{tr}(\overline{\sigma}_{i,i}) \left[ S(\overline{R}_{i,i}( \overline{\sigma}_{i,i} / \mathrm{tr}(\overline{\sigma}_{i,i}))) - S( \overline{\sigma}_{i,i} / \mathrm{tr}(\overline{\sigma}_{i,i}) ) \right],
\end{equation}
 where we sample computational basis states of subsystem $B'$ from the distribution $\mathrm{tr}(\overline{\sigma}_{i,i})$
 and compute the entropy production from the normalized $\overline{\sigma}_{i,i}$.
The computational cost of each sample grows exponentially with the number of elementary subsystems in $\overline{B'}$.

Because of the Hermitian positive semi-definite structure of $R_{ij,kl}$ in Eq.\@ (\ref{general_R}),
 it is not straightforward to apply unconstrained machine-learning models to the general quantum case.
The most natural variational form is to decompose $R$ into a sequence of operations $R^{(i)}$, each acting on a different small cluster of elementary subsystems $\overline{B'_i}$.
We begin with a simple extension to subsystem $A$, $\sigma^{(0)} = |0\rangle \langle 0| \otimes \sigma$,
 and form $\rho = \sigma^{(n)}$ as a sequence of $n$ operations, $\sigma^{(i+1)} = R^{(i)}(\sigma^{(i)})$.
Each $R^{(i)}$ has the general form
\begin{equation} \label{unextending_R}
 R^{(m)}(\rho_{A} \otimes \rho_{\overline{B'_m}} \otimes \rho_{B'_m}) = \sum_{i,j = 0}^{N_A^2-1} \sum_{k,l = 0}^{N_{B'_m}-1} R^{(m)}_{ik,jl} \, C_i \rho_{A} C_j^\dag \otimes |k\rangle \langle k | \rho_{\overline{B'_m}}  | l \rangle \langle l | \otimes \rho_{B'_m}
\end{equation}
 for some operator basis $C_i$ on subsystem $A$.
Hermitian positive semi-definite structure is still required, but for $n$ small $R^{(m)}_{ik,jl}$ matrices rather than a large $R_{ik,jl}$ matrix.
There are many ways to enforce Hermitian positive semi-definite structure on small, dense matrices
 such as storing a matrix $F$ in a factored form, $F = G G^\dag$, for an upper-triangular matrix $G$.
This cluster expansion of $R$ allows for separate calculations of entropy production for each $R^{(i)}$
 using a different choice of $B'$ to define $\sigma_0$ in Eq.\@ (\ref{repartition}) for each calculation,
 which allows us to balance variational freedom and tightness of bounds with $B' = B'_i$ and
\begin{equation}\label{quantum_entropy_bound3}
 S(\sigma^{(i+1)}) \ge S(\sigma^{(i)}) + \sum_{j=0}^{N_{B'_i}-1}  \mathrm{tr}(\overline{\sigma}^{(i)}_{j,j}) \left[ S( \overline{R}^{(i)}(\overline{\sigma}^{(i)}_{j,j} / \mathrm{tr}(\overline{\sigma}^{(i)}_{j,j}))) - S( \overline{\sigma}^{(i)}_{j,j} / \mathrm{tr}(\overline{\sigma}^{(i)}_{j,j}) ) \right].
\end{equation}
The partitioned form of $R^{(i)}$ in Eq.\@ (\ref{R_repartition}) does not have subsystem dependence, $\overline{R}^{(i)}_{jk} = \overline{R}^{(i)}$,
 since it does not act on $B'_i$ in Eq.\@ (\ref{unextending_R}).
Using a cluster expansion of $R$ only increases the non-uniqueness of the robust decomposition,
 but the choice of decomposition has an effect on the tightness of entropy bounds
 and free-energy minimization might produce a unique decomposition that optimizes their tightness.
For example, the decomposition using $\sigma$ in Eq.\@ (\ref{rank1_sigma})
 defers all entropy production until the final subsystem is prepared,
 which causes entropy bounds to be tight for all but the final $R$
 but requires the preparation of a single subsystem to generate
 all of the entropy in $\rho$.
It is likely that the tightest overall entropy bound
 will result from distributing entropy production more evenly over every $R$.

\subsection{Conditional hybrid decomposition}

Compatibility between unconstrained machine-learning models and robust decompositions of mixed quantum states
 can be improved by introducing ancillary subsystems.
The most popular approach is state purification, whereby a mixed state on subsystem $A$,
\begin{equation}
 \rho_A = \sum_{i=0}^{N_A - 1} p_i |\psi_i \rangle \langle \psi_i |,
\end{equation}
 is rewritten as the partial trace of a pure state on subsystems $A$ and $B$,
\begin{equation}
 \rho_A = \mathrm{tr}_B | \Psi_{AB} \rangle \langle \Psi_{AB} | \ \ \ \mathrm{for} \ \ \ | \Psi_{AB} \rangle = \sum_{i=0}^{N_A - 1} \sqrt{p_i} |\psi_i \rangle \otimes | i \rangle,
\end{equation}
 assuming that subsystem $B$ has the same dimensionality as $A$.
While ancillary degrees of freedom can introduce even more indeterminacy into robust decompositions,
 they are also beneficial in many ways.
For example, the von Neumann entropy of $\rho_A$ is equivalent to an entanglement entropy of $|\Psi_{AB}\rangle$,
 which can be calculated using Monte Carlo methods \cite{Renyi_entanglement}.
Here we discuss two distinct ancilla-based methods for calculating von Neumann entropy
 using robust decompositions built from unconstrained machine-learning models.

The first method is to introduce two temporary classical ancillary subsystems with the same dimensionality as $A$ and $B$
 that are used to construct $R_{ij,kl}$ in Eq.\@ (\ref{general_R}) as
\begin{align} \label{machine_learning_R}
 R_{ij,kl} &= \sum_{m=0}^{N_B - 1} P_B(m) R_{ij,kl}^{(m)} \ \ \ \mathrm{for} \ \ \ \notag \\
 R_{ij,kl}^{(n)} &= \sum_{m=0}^{N_A-1} P_{A|B}(m | n) \psi_{A|B}(i|j;m,n) \psi_{A|B}(k|l;m,n)^*,
\end{align}
 from both a sequence of conditional probabilities $P_{A|B}$ as discussed in Sec.\@ \ref{probability_decomposition_subsection}
 and a set of conditional quantum states $\psi_{A|B}$ as discussed in Sec.\@ \ref{quantum_decomposition_subsection}
 that are parameterized by the temporary classical variables.
We can then use machine-learning models to construct both $P_{A|B}$ and $\psi_{A|B}$.
While we can no longer efficiently compute the entropy lower bounds in Eq.\@ (\ref{quantum_entropy_bound1}), we can still efficiently compute
 a more relaxed form of the bound,
\begin{equation} \label{quantum_entropy_bound4}
 S(\rho) \ge S(\sigma) + \sum_{i,j=0}^{N_B-1} \langle i | \sigma | i \rangle P_B(j) S(R^{(j)}(|i\rangle \langle i|)) ,
\end{equation}
 by also sampling from the classical variables in the temporary version of subsystem $B$ in addition to
 sampling from the computational basis in subsystem $B$.
This bound can still be systematically improved until it is tight using the strategies suggested in Sec.\@ \ref{density_decomposition_subsection}.
We can expand the subsystem over which $\sigma_0$ and $\sigma$ have matching marginal states
 and at the same time marginalize over the corresponding temporary classical variables
 so that they are moved from the sum in Eq.\@ (\ref{quantum_entropy_bound4}) to the definition of $R^{(n)}$ in Eq.\@ (\ref{machine_learning_R}).
Similarly, we can also decompose $R$ into a sequence of operations as in Eq.\@ (\ref{unextending_R})
 and use a different $\sigma_0$ and set of marginalized temporary classical variables to calculate lower bounds
 on the entropy production of each operation in the sequence, analogous to Eq.\@ (\ref{quantum_entropy_bound3}).

The second method is to introduce two permanent classical ancillary subsystems with the same dimensionality as $A$ and $B$
 that is used to define a mixed quantum state as
\begin{align} \label{hidden_ancilla}
 \rho &= \sum_{i=0}^{N_A-1} \sum_{j=0}^{N_B-1} P_{AB}(i, j) | \psi_{AB}(i,j) \rangle \langle \psi_{AB}(i,j)| \ \ \ \mathrm{for} \notag \\
 | \psi_{AB}(i,j) \rangle &= \sum_{k=0}^{N_A-1} \sum_{l=0}^{N_B-1} \psi_{AB}(k,l;i,j) |k\rangle \otimes |l\rangle,
\end{align}
 where again both $P_{AB}$ and $\psi_{AB}$ are defined recursively using conditional decompositions $P_{A|B}$ and $\psi_{A|B}$
 that are compatible with unconstrained machine-learning models.
Unlike the other states discussed in Sec.\@ \ref{decomposition_section}, this state only has a convenient robust decomposition
 when the classical ancilla is also part of the reconstruction operation, $\rho_+ = R_+(\sigma_+)$ for
\begin{align}
 \rho_+ &= \sum_{i=0}^{N_A - 1}  \sum_{j=0}^{N_B - 1} P_{AB}(i, j) | \psi_{AB}(i,j) \rangle \langle \psi_{AB}(i,j) | \otimes |i\rangle \langle i| \otimes |j\rangle \langle j| , \notag \\
 \sigma_+ &= \sum_{i=0}^{N_B - 1} P_{B}(i) | \psi_{B}(i) \rangle \langle \psi_{B}(i) | \otimes |i\rangle \langle i| , \notag \\
 R_+(\rho_{B} \otimes \sigma_{B} ) &= \sum_{i=0}^{N_A-1} \sum_{j=0}^{N_B-1} P_{A|B}(i | j) V(i,j) \rho_B V(i,j)^\dag \otimes |i \rangle \langle i | \otimes |j\rangle \langle j| \sigma_B  |j\rangle \langle j| , \notag \\
 V(i,j) &= \sum_{k=0}^{N_A-1} \sum_{l=0}^{N_B-1} \psi_{A|B}(k | l ; i,j ) | k \rangle \otimes |l\rangle \langle l| ,
\end{align}
 which can be reversed by the ancilla-dependent marginalization operation
\begin{equation}
 M_+ (\rho_{AB} \otimes \sigma_A \otimes \sigma_B ) = \sum_{i=0}^{N_A-1} \sum_{j=0}^{N_B-1} V(i,j)^\dag \rho_{AB} V(i,j) \otimes \langle i | \sigma_A | i \rangle \otimes |j\rangle \langle j| \sigma_B |j\rangle \langle j| .
\end{equation}
Although the von Neumann entropy of $\rho_+$ is equal to the efficiently computable Shannon entropy of $P_{AB}$,
 it is an upper bound on the entropy of $\rho$ that is not variationally stable --
 minimization of an approximate free-energy calculated from this entropy bound will drive all $| \psi_{AB}(i,j) \rangle$
 towards the ground state and $P_{AB}$ towards the maximally mixed state.

To construct an entropy lower bound for $\rho$ in Eq.\@ (\ref{hidden_ancilla}),
 we again use the monotonicity of quantum relative entropy in Eq.\@ (\ref{monotonicity})
 with a different choice of states and operations,
\begin{equation} \label{monotonicity2}
 S( \rho_+ \| \rho \otimes I_{AB} ) \ge S( J(\rho_+) \| J(\rho \otimes I_{AB} ) ),
\end{equation}
 where $J$ performs quantum measurements on the non-ancillary subsystems and classical post-processing
 to recover classical labels $(i,j)$ of states $| \psi_{AB}(i,j) \rangle$ and $I_{AB}$ is the identity operator on the ancillary subsystems.
We can rearrange Eq.\@ (\ref{monotonicity2}) into the lower bound
\begin{equation}
 S( \rho ) \ge S( P_{AB} ) - S( J(\rho_+) ) + S(J(\rho \otimes I_{AB} )),
\end{equation}
 where the right-hand side can be interpreted as the classical mutual information between the ancillary subsystems
 and the distribution of classical labels produced by $J$.
This is a single-step bound on $S( \rho )$ that requires $J$ to extract a lot of information from $\rho$ whereas previous bounds were more incremental.
We can decompose this single-step bound into a recursive sequence of bounds by defining an intermediate state between $\rho$ and $\rho_+$,
\begin{equation}
 \rho_* = \sum_{i=0}^{N_A - 1}  \sum_{j=0}^{N_B - 1} P_{AB}(i, j) | \psi_{AB}(i,j) \rangle \langle \psi_{AB}(i,j) | \otimes I \otimes |j\rangle \langle j|,
\end{equation}
 and a more limited operation $K$ that is only required to recover the classical labels from ancillary subsystem $A$
  while being able to measure both the non-ancillary subsystems and ancillary $B$ subsystem.
The more limited version of Eq.\@ (\ref{monotonicity2}) is
\begin{equation} \label{monotonicity3}
 S( \rho_+ \| \rho_* ) \ge S( K(\rho_+) \| K(\rho_* ) ),
\end{equation}
 and the corresponding lower bound on the entropy of $\rho_*$ is
\begin{equation}
 S( \rho_* ) \ge S(P_B) + \sum_{i=0}^{N_B - 1} P_B(i) \left[ S( P_{AB} ) - S( K(\rho_+) ) + S( K(\rho_* ) ) \right]_{B = i} ,
\end{equation}
 where $[\cdots]_{B=i}$ denotes the conditional states of $P_{AB}$, $K(\rho_+)$, and $K(\rho_*)$
 with the ancillary $B$ subsystem in state $i$.
This corresponds to the classical conditional mutual information between the ancillary subsystem $A$
 and the distribution of classical labels produced by $K$ being added to the entropy of $P_B$.
The recursive process continues with $\rho_*$ instead of $\rho_+$ by decomposing the ancillary subsystem $B$
 into two more subsystems and recovering the classical label from the smaller subsystem.
The implementation details of $K$ are flexible -- optimal distinguishing measurements are known for some situations \cite{state_discrimination},
 but any practical positive operator-valued measure (POVM) must have low quantum weight to be efficiently observable within our noisy quantum circuit model.

While machine-learning models can be more flexible variational forms than the cluster decompositions
 discussed in Sec.\@ \ref{density_decomposition_subsection}, the tightness of their entropy lower bounds is limited
 by our restricted ability to observe states rather than restricted variational freedom.
The cluster decompositions attempt to balance these two restrictions, while machine-learning models
 provide an opportunity to saturate variational freedom and eliminate that source of error.
Machine-learning models also enable two distinct forms of entropy lower bounds,
 one limited by reversibility of the reconstruction process and another that has a perfectly reversible reconstruction process
 but is limited by the distinguishability of its underlying pure states $|\psi_{AB}(i,j)\rangle$ in Eq.\@ (\ref{hidden_ancilla}).
The practical efficacy of these different approaches is not obvious from our preliminary theoretical analysis,
 and they will have to be compared in practical tests of their viability as finite-temperature VMC methods.

\section{Numerical examples}

We demonstrate the viability of a finite-temperature VMC method with some inaugural numerics on the one-dimensional Ising model.
This well-understood example simplifies the construction of a minimum viable numerical method by exploiting translational invariance
 and the locality of correlations in some parameter regimes.
We restrict the reconstruction operations $R$ from Sec.\@ \ref{decomposition_section}
 to be dependent only on the $n$ spins immediately preceding each newly prepared spin
 and each new $R$ is a single-site translation of the previous $R$.
Entropy calculations are similarly restricted to surviving clusters of $n+1$ contiguous spins
 for use as reference states $\sigma_0$ in Eqs.\@ (\ref{repartition}) and (\ref{quantum_entropy_bound2}).
With a translationally invariant system, we can also use spatial averaging as a convenient substitute for sample averaging,
 although this will correlate observations within a spatial correlation length and reduce the efficiency
 of the sampling process.
We also track how the accuracy of these calculations changes as a function of correlation length.
These results may inform further analysis of the various methodological design choices proposed in Sec.\@ \ref{decomposition_section}
 despite not testing them directly.

The numerical implementation used here is designed for simplicity over efficiency \cite{github_repo}.
We sample from states of surviving spin clusters that are conditioned on computational basis measurements of all the other spins
 that are not contained in a surviving spin cluster.
This sampling process is split into generating a sequence of measurement outcomes
 and constructing the conditional states associated with those measurements.
Many outcomes at the beginning of the sampling process are discarded to allow for spatial equilibration of the sampled spin distribution.
Derivatives of the sample free energy are calculated for a fixed set of measurement outcomes
 to generate a search direction in the model parameter space.
The search direction is the gradient preconditioned by the diagonals of the Hessian matrix,
 and it defines a line search over which the free energy is minimized.
The sampling process uses the same pseudorandom sequence each time,
 which causes the sampled free energy to be deterministic but increasingly discontinuous in the number of samples.
This implementation maintains non-trivial derivatives of the sampled free energy with respect to all model parameters,
 which avoids the problem of barren plateaus in the optimization landscape \cite{barren_plateau}
 by carefully pairing model parameters with observations.
Also, it allows us to use numerical derivatives of the sampled free energy for simplicity,
 although it may be more effective to sample analytical derivatives directly \cite{stochastic_gradients}
 as implementations mature.

\subsection{Classical example}

The Hamiltonian for the classical one-dimensional Ising model is
\begin{equation} \label{classical_ising}
 H = \alpha \sum_i Z_i - \sum_i Z_i Z_{i+1} ,
\end{equation}
 where $\alpha$ is the magnetic field strength and $Z_i$ is the conventional Pauli operator for spin $i$.
Conditional probabilities that do not depend on previous spins correspond to a mean-field theory,
 which parameterizes the free energy per site as
\begin{equation}
 F(p) = \alpha (1 - 2 p) - (1 - 2 p)^2 + T p \ln (p) + T (1-p) \ln (1-p),
\end{equation}
 where $p$ is the probability for a spin to be in state $|0\rangle$ and $T$ is the temperature.
This free energy is then minimized over $p \in [0,1]$, which causes an erroneous phase transition when
 a varying minimum for $p \in (0,1)$ saturates to $p \in \{0,1\}$ at low temperatures.
Conditional probabilities that depend on one previous spin are capable of representing exact solutions
 because the Boltzmann distribution, $e^{-H/T}/ \mathrm{tr}(e^{-H/T})$, can be factored into two-variable
 functions that are normalized as conditional probability functions,
\begin{align}
 \frac{\langle \mathbf{s} | e^{-H/T} | \mathbf{s} \rangle}{\mathrm{tr}(e^{-H/T})} &= \prod_{i} P(s_{i+1} | s_{i}) \ \ \ \mathrm{for} \notag \\
 P(i|j) &= \frac{e^{-[\alpha - (1-2i)] (1 - 2j)/T}}{\sum_{k=0}^{1} e^{-[\alpha - (1-2k)] (1 - 2j)/T}} \frac{\sum_{k=0}^{1} e^{-[\alpha - (1-2i)] (1 - 2k)/T}} {\sum_{k=0}^{1} \sum_{l=0}^{1} e^{-[\alpha - (1-2k)] (1 - 2l)/T}} ,
\end{align}
 where $s_i$ are the individual spin configurations within the computational basis vector $|\mathbf{s}\rangle$.
Thus, a nearest-neighbor conditional dependence is able to generate the exact long-range spin correlations
 at arbitrarily long distances, and there is no direct relationship between the length scales of correlations and conditional dependencies.

\begin{figure}
\centering
\includegraphics{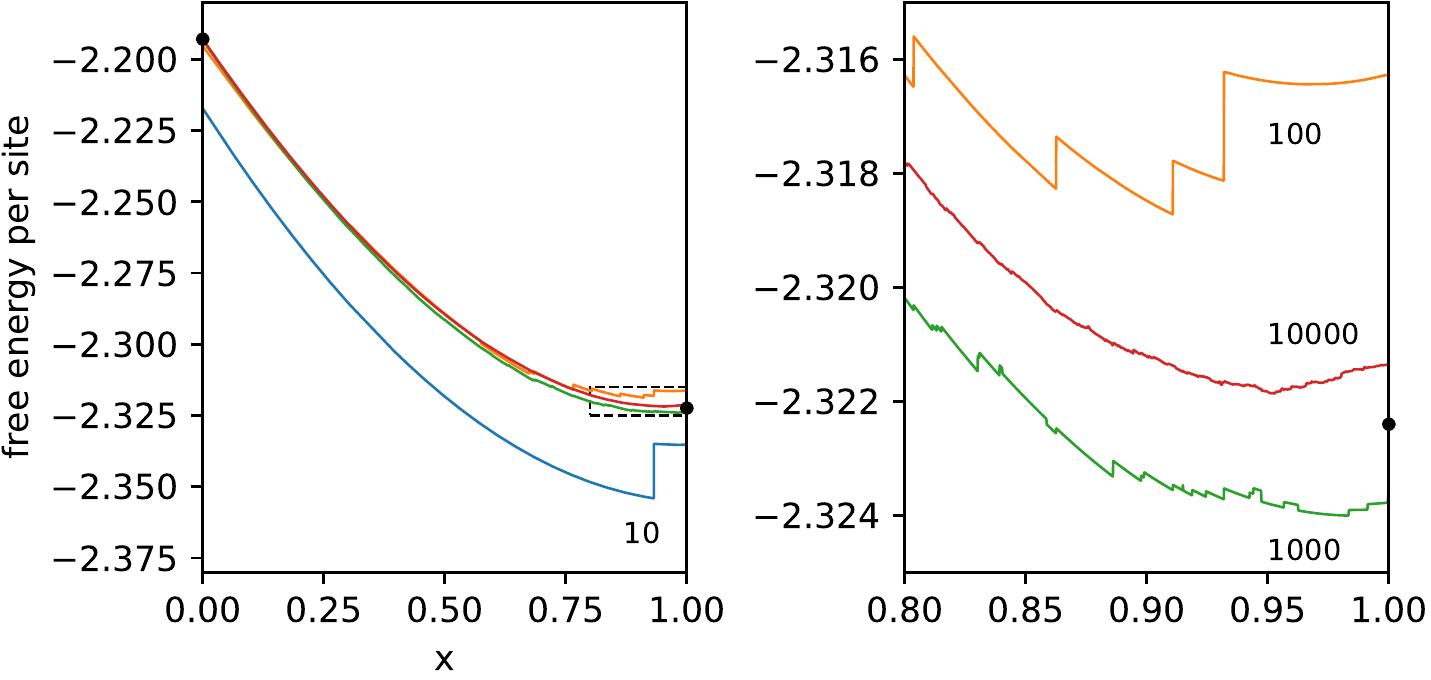}
\caption{\label{energy_line_search} A line search over free energy between the mean-field ($x=0$) and exact solution ($x=1$)
of Eq.\@ (\ref{classical_ising}) for $\alpha = 0.5$ and $T=3$ with varying numbers of pseudorandom samples (as labeled).}
\end{figure}

While this example is not a stringent test of our variational ansatz, it can demonstrate some features of our numerical implementation.
In Fig.\@ \ref{energy_line_search}, we illustrate the behavior of a deterministic but pseudorandom free-energy surface
 that becomes more accurate with an increasing number of samples but also more discontinuous as the number of measurement outcomes increases.
The local curvature of the free-energy surface is representative of the numerical derivative values at fixed measurement outcomes.
These derivatives are useful for defining search directions, but the free-energy minimizations along a search direction cannot make strong assumptions about continuity.
The sampled free energy is also not an upper bound on the minimum free energy, unlike the variational free energy in the limit of infinite samples.
If the standard error in the sampled free energy becomes larger than the variational error,
 it could trap the free energy minimization in an erroneously low free energy with a large standard error.
In these circumstances, it might be more reliable to minimize the sampled free energy plus a multiple of the standard error,
 which functions as a looser but more statistically certain upper bound on free energy.

\subsection{Pure-state example\label{pure_example}}

We introduce quantum effects in the one-dimensional Ising model
 by replacing the parallel magnetic field in Eq.\@ (\ref{classical_ising})
 with a transverse magnetic field,
\begin{equation} \label{quantum_ising}
 H = \alpha \sum_i X_i - \sum_i Z_i Z_{i+1}.
\end{equation}
This is a well-studied Hamiltonian with an exact analytical solution \cite{transverse_1D_Ising}
 constructed by a Jordan-Wigner transformation from interacting spins to non-interacting fermions.
This is a convenient source of reference data with a complicated enough mathematical structure
 that our variational ansatze are not exact, unlike the classical case.
To give useful context to small numerical errors, we compare the ground-state energy per site with second-order perturbation theory
 from either the low-field ($\alpha \ll 1$) or high-field ($\alpha \gg 1$) limits,
\begin{equation} \label{second_order}
 E / N \approx \left\{ \begin{array}{ll} -1 - \alpha^2 /4 , & \alpha < 1 \\ -\alpha - \alpha^{-1} /4 , & \alpha \ge 1 \end{array} \right. .
\end{equation}
This simple result captures all but a few percent of the energy
 with a maximum error at $\alpha = 1$, the quantum critical point of the Hamiltonian.
It is an estimate for the amount of ground-state energy that can be recovered from simple calculations.

\begin{figure}
\centering
\includegraphics{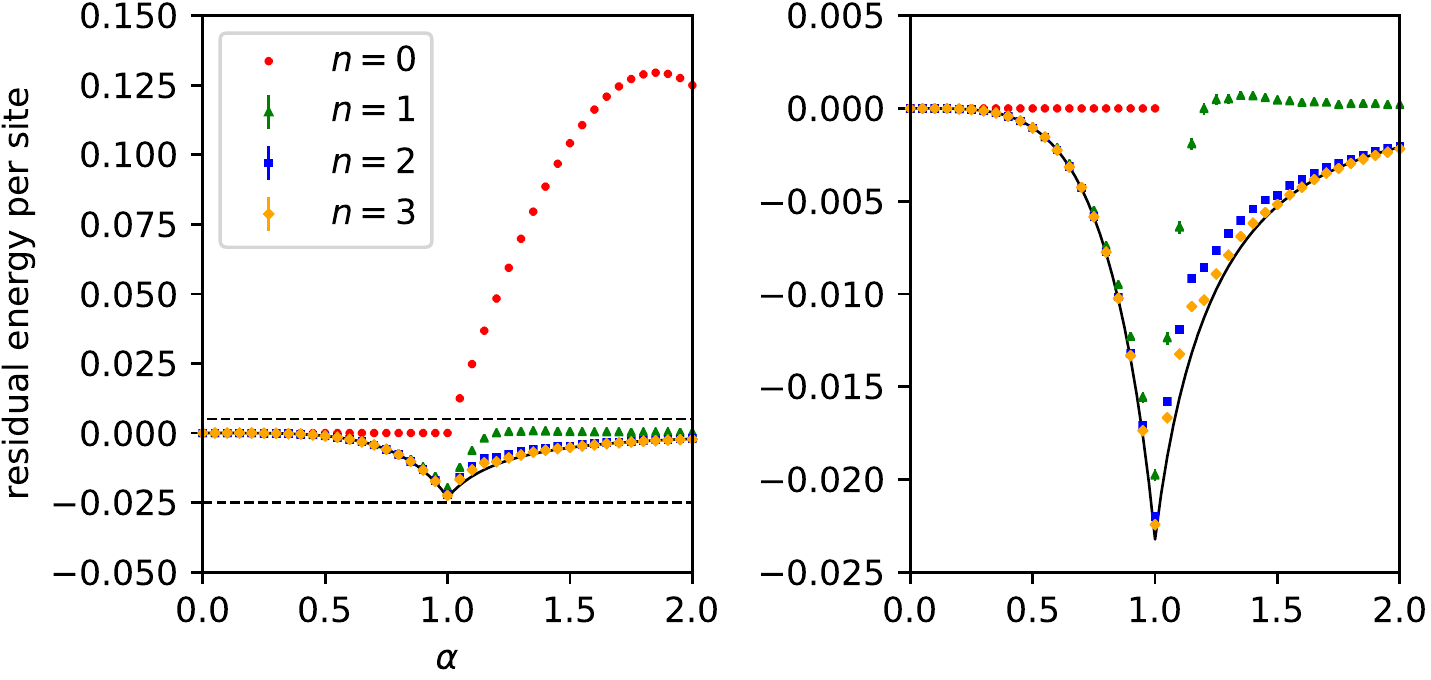}
\caption{\label{pure_energy_scan} Ground-state energy of Eq.\@ (\ref{quantum_ising}) relative to Eq.\@ (\ref{second_order}).
VMC calculations in which qubits are conditioned on $n$ previous qubits are compared to the exact result (solid line).
}
\end{figure}

Our numerical results are shown in Fig.\@ \ref{pure_energy_scan}.
In each case, we sample $10^5$ sites from the infinite lattice
 and discard the first $10^3$ sites to avoid burn-in effects.
We observed some instances of energy minimization getting trapped at local minima,
 which we avoided by initializing states conditioned on $n$ qubits
 with the minimum-energy state conditioned on $n-1$ qubits at the same $\alpha$ value.
As in the classical case, $n=0$ corresponds to mean-field theory, which we can calculate exactly without statistical sampling.
Mean-field errors get as large as $\approx 5\%$ of the overall energy scale of the Hamiltonian
 while the second-order perturbation theory errors remain less than $\approx 1 \%$.
Variational errors are largest near the critical point at $\alpha =1$,
 but the local variational ansatz is surprisingly accurate at $\alpha = 1$,
 which has long-range algebraically decaying pair correlations in its ground state.
Just as with classical correlations, a local state reconstruction process
 can propagate long-range quantum entanglement between well-separated qubits.
At $n=3$, the largest variational errors are less than $\approx 0.1 \%$ of the overall energy scale of the Hamiltonian.

\subsection{Mixed-state example}

Free energy minimization of Eq.\@ (\ref{quantum_ising}) at a finite temperature $T$ has many similarities with energy minimization at zero temperature.
Its exact solution extends to finite temperature because non-interacting fermions remain exactly solvable at finite temperature.
We can again use second-order perturbation theory to calculate the free energy per site from the high-field limit that extends the $T=0$ and $\alpha \ge 1$ result from Eq.\@ (\ref{second_order}) to
\begin{equation} \label{second_order_thermal}
 F / N \approx - T \ln (2 \cosh(\alpha/T)) - \frac{2 \alpha + T \sinh(2 \alpha / T)}{8\alpha T \cosh^2(\alpha / T)} .
\end{equation}
We again use this as the point of reference for numerical errors, which captures all but a few percent of the free energy.
Although high temperatures suppress quantum effects and make simulations easier,
 the errors in this approximation remain appreciable while $T$ is less than or proportional to the energy scale of the Hamiltonian.
High-temperature expansion is another common approximation method, and it diverges in this temperature regime.

\begin{figure}
\centering
\includegraphics{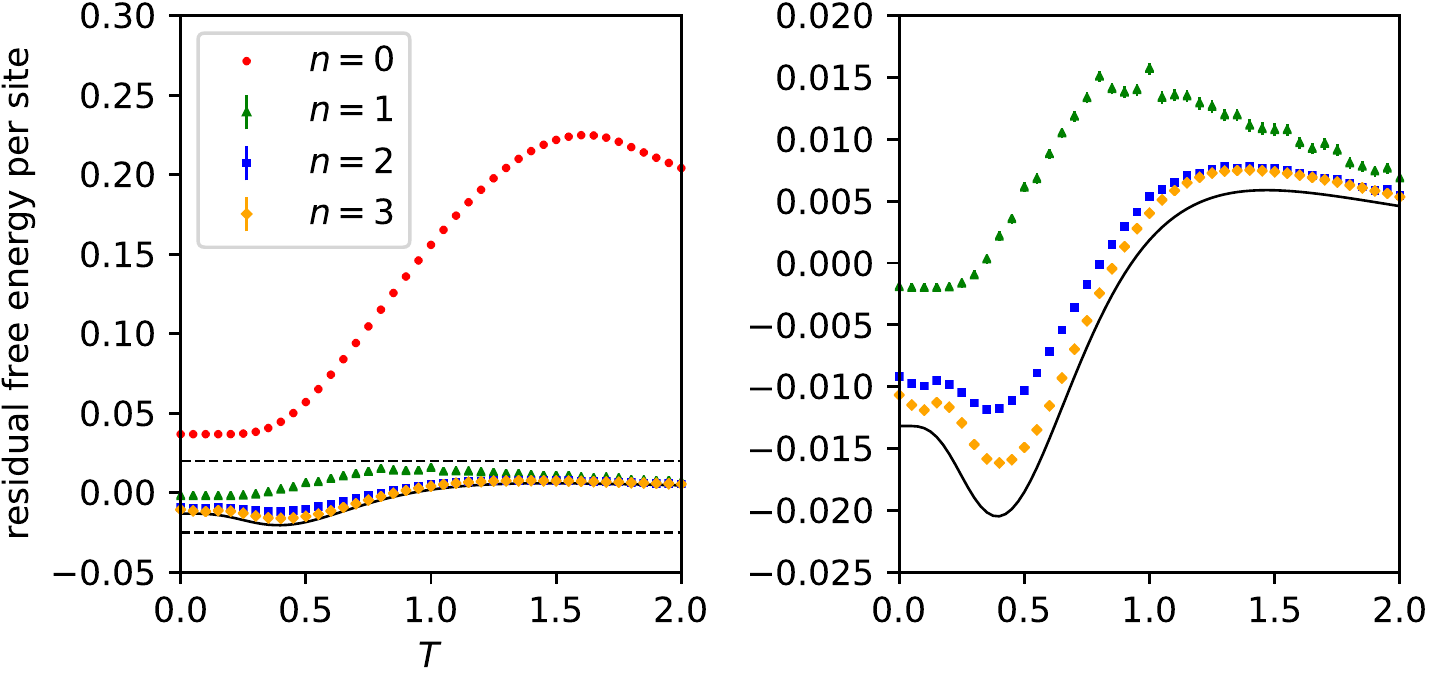}
\caption{\label{mixed_energy_scan} Free energy of Eq.\@ (\ref{quantum_ising}) at $\alpha = 1.15$ relative to Eq.\@ (\ref{second_order_thermal}).
VMC calculations in which qubits are conditioned on $n$ previous qubits are compared to the exact result (solid line).}
\end{figure}

Our finite-temperature numerical results are shown in Fig.\@ \ref{mixed_energy_scan}.
We focus on the field value of $\alpha = 1.15$, which has the largest variational errors at $T = 0$.
These errors clearly persist at intermediate temperatures, and only at high temperatures ($T>2$) do all of the approximations
 coalesce onto a common high-temperature limit.
The mean-field solution ($n=0$) has larger errors at intermediate temperatures than $T=0$,
 and it approaches the high-temperature limit much more slowly than other solutions.
Some finite-sampling artifacts are visible in the $n=1$ solutions at intermediate temperatures where variance is largest.
Variance is reduced with larger $T$ and $n$ because statistical variations within the surviving qubit clusters
 are calculated exactly without sampling variance.
We tested the tightness of entropy lower bounds indirectly in these calculations by recalculating entropy bounds with one fewer qubits dephased in the reference state in Eq.\@ (\ref{quantum_entropy_bound2}).
The calculated entropy values did not increase noticeably, which suggests that variational errors are much larger than entropy errors resulting from loose lower bounds.
At $n=3$, variational errors remain less than $\approx 0.2 \%$ of the overall energy scale of the Hamiltonian.
The calculations at $T>0$ remain nearly as accurate as the $T=0$ calculations in Sec.\@ \ref{pure_example}, thus we have demonstrated a practical finite-temperature VMC method.

\section{Conclusions}

In summary, this paper has presented a theoretical framework that manifests the common belief that noise and uncertainty reduce the classical simulation cost of quantum systems.
It establishes a regime of classical-quantum computational equivalence in which the noise rates
 of a specific noise model are able to control the classical simulation costs of quantum systems
 with a smooth crossover from a low-noise regime in which quantum systems are more efficient
 to a high-noise regime in which classical computers are more efficient.
This result has no effect on the long-term prospects of UQC and the profound computational advantages 
 of quantum algorithms like Grover's algorithm and Shor's algorithm and the ability to perform unbiased quantum simulations.
However, the modest numerical results presented here demonstrate that our theoretical framework can be adapted into practical
 finite-temperature VMC methods.
The methods operate on hypothetical, idealized NISQ devices with a limited set of heralded errors,
 whereas variational quantum eigensolvers \cite{VQE} must operate on physical NISQ devices
 and accommodate device-specific limitations and physical noise while constructing problem-specific variational forms.
This frustrates both the design and optimization landscape of their variational forms \cite{barren_plateau}.
As they develop and mature, these new finite-temperature VMC methods may become highly competitive with physical NISQ devices in applications of biased quantum simulation.

\begin{acknowledgements}
The Molecular Sciences Software Institute is supported by grant ACI-1547580 from the National Science Foundation.
\end{acknowledgements}


\begin{thebibliography}{99}
\bibitem{NISQ} J. Preskill, \href{https://doi.org/10.22331/q-2018-08-06-79}{Quantum \textbf{2}, 79 (2018)}.
\bibitem{Google_supremacy} F. Arute, K. Arya, R. Babbush, D. Bacon, J. C. Bardin, R. Barends, R. Biswas, S. Boixo, F. G. S. L. Brandao, D. A. Buell, B. Burkett, Y. Chen, Z. Chen, B. Chiaro, 
R. Collins, W. Courtney, A. Dunsworth, E. Farhi, B. Foxen, A. Fowler, C. Gidney, M. Giustina, R. Graff, K. Guerin, S. Habegger, M. P. Harrigan, M. J. Hartmann, A. Ho, M. Hoffmann, T. Huang,
T. S. Humble, S. V. Isakov, E. Jeffrey, Z. Jiang, D. Kafri, K. Kechedzhi, J. Kelly, P. V. Klimov, S. Knysh, A. Korotkov, F. Kostritsa, D. Landhuis, M. Lindmark, E. Lucero, D. Lyakh, S. Mandr\`{a},
J. R. McClean, M. McEwen, A. Megrant, X. Mi, K. Michielsen, M. Mohseni, J. Mutus, O. Naaman, M. Neeley, C. Neill, M. Y. Niu, E. Ostby, A. Petukhov, J. C. Platt, C. Quintana, E. G. Rieffel,
P. Roushan, N. C. Rubin, D. Sank, K. J. Satzinger, V. Smelyanskiy, K. J. Sung, M. D. Trevithick, A. Vainsencher, B. Villalonga, T. White, Z. J. Yao, P. Yeh, A. Zalcman, H. Neven, and
J. M. Martinis, \href{https://doi.org/10.1038/s41586-019-1666-5}{Nature \textbf{574}, 505 (2019)}.
\bibitem{VMC_origin} W. L. McMillan, \href{https://doi.org/10.1103/PhysRev.138.A442}{Phys. Rev. \textbf{138}, A442 (1965)}.
\bibitem{QAOA} E. Farhi, J. Goldstone,  and S. Gutmann, \href{https://arxiv.org/abs/1411.4028}{arXiv:1411.4028}.
\bibitem{VQE} A. Peruzzo, J. McClean, P. Shadbolt,  M.-H. Yung, X.-Q. Zhou, P. J. Love, A. Aspuru-Guzik, and J. L. O’Brien, \href{https://doi.org/10.1038/ncomms5213}{Nat. Commun. \textbf{5}, 4213 (2014)}.
\bibitem{QPE} D. S. Abrams and S. Lloyd, \href{https://doi.org/10.1103/PhysRevLett.83.5162}{Phys. Rev. Lett. \textbf{83}, 5162 (1999)}.
\bibitem{NISQ_complexity} A. Bouland, B. Fefferman, C. Nirkhe, and U. Vazirani, \href{https://doi.org/10.1038/s41567-018-0318-2}{Nat. Phys. \textbf{15},159 (2019)}.
\bibitem{noise_dilution} I. L. Markov, A. Fatima, S. V. Isakov, and S. Boixo, \href{https://arxiv.org/abs/1807.10749}{arXiv:1807.10749}.
\bibitem{stabilizer_QEC} D. Gottesman, \href{https://arxiv.org/abs/quant-ph/9705052}{arXiv:quant-ph/9705052}.
\bibitem{stabilizer_simulation} S. Aaronson and D. Gottesman, \href{https://doi.org/10.1103/PhysRevA.70.052328}{Phys. Rev. A \textbf{70}, 052328 (2004)}.
\bibitem{magic_state_distillation} S. Bravyi and A. Kitaev, \href{https://doi.org/10.1103/PhysRevA.71.022316}{Phys. Rev. A \textbf{71}, 022316 (2005)}.
\bibitem{random_sampling} S. Boixo, S. V. Isakov, V. N. Smelyanskiy, R. Babbush, N. Ding, Z. Jiang, M. J. Bremner, J. M. Martinis, and H. Neven, \href{https://doi.org/10.1038/s41567-018-0124-x}{Nature Phys. \textbf{14}, 595 (2018)}.
\bibitem{IQP_sampling} M. J. Bremner , R. Jozsa, and D. J. Shepherd, \href{https://doi.org/10.1098/rspa.2010.0301}{Proc. R. Soc. A \textbf{467}, 459 (2011)}.
\bibitem{boson_sampling} S. Aaronson and A. Arkhipov, \href{https://doi.org/10.4086/toc.2013.v009a004}{Theory Comput. \textbf{9}, 143 (2013)}.
\bibitem{supremacy_verification} D. Hangleiter, M. Kliesch, J. Eisert, and C. Gogolin, \href{https://doi.org/10.1103/PhysRevLett.122.210502}{Phys. Rev. Lett. \textbf{122}, 210502 (2019)}.
\bibitem{boson_noise} R. Garc\'{i}a-Patr\'{o}n, J. J. Renema, and V. Shchesnovich, \href{https://doi.org/10.22331/q-2019-08-05-169}{Quantum \textbf{3}, 169 (2019)}.
\bibitem{IQP_noise} M. J. Bremner, A. Montanaro, and D. J. Shepherd, \href{https://doi.org/10.22331/q-2017-04-25-8}{Quantum \textbf{1}, 8 (2017)}.
\bibitem{bond_dimension} J. Haegeman, C. Lubich, I. Oseledets, B. Vandereycken, and F. Verstraete, \href{https://doi.org/10.1103/PhysRevB.94.165116}{Phys. Rev. B \textbf{94}, 165116 (2016)}.
\bibitem{quasiprobability} H. Pashayan, J. J. Wallman, and S. D. Bartlett, \href{https://doi.org/10.1103/PhysRevLett.115.070501}{Phys. Rev. Lett. \textbf{115}, 070501 (2015)}.
\bibitem{robust_state} G. Vidal and R. Tarrach, \href{https://doi.org/10.1103/PhysRevA.59.141}{Phys. Rev. A \textbf{59}, 141 (1999)}.
\bibitem{robust_gate} A. W. Harrow and M. A. Nielsen, \href{https://doi.org/10.1103/PhysRevA.68.012308}{Phys. Rev. A \textbf{68}, 012308 (2003)}.
\bibitem{robust_operation} K. M{\o}lmer and A. S{\o}rensen, \href{https://doi.org/10.1103/PhysRevLett.82.1835}{Phys. Rev. Lett. \textbf{82}, 1835 (1999)}.
\bibitem{robust_control} D. Dong and I. R. Petersen, \href{doi: 10.1049/iet-cta.2009.0508}{IET Control Theory Appl. \textbf{4}, 2651 (2010)}.
\bibitem{robust_control_book} K. Zhou and J. Doyle, \textit{Essentials of Robust Control} (Prentice Hall, New Jersey, 1998).
\bibitem{robust_polynomial} H. Buhrman, I. Newman, H. Rohrig, and R. de Wolf, \href{https://doi.org/10.1007/s00224-006-1313-z}{Theory Comput. Syst. \textbf{40}, 379 (2007)}.
\bibitem{robust_learning} A. W. Cross, G. Smith, and J. A. Smolin, \href{https://doi.org/10.1103/PhysRevA.92.012327}{Phys. Rev. A \textbf{92}, 012327 (2015)}.
\bibitem{robust_simulation} P. Hauke, F. M. Cucchietti, L. Tagliacozzo, I. Deutsch, and M. Lewenstein, \href{https://doi.org/10.1088/0034-4885/75/8/082401}{Rep. Prog. Phys. \textbf{75}, 082401 (2012)}.
\bibitem{FTQEC} D. P. DiVincenzo and P. W. Shor, \href{https://doi.org/10.1103/PhysRevLett.77.3260}{Phys. Rev. Lett. \textbf{77}, 3260 (1996)}.
\bibitem{quantum_bayesian} R. R. Tucci, \href{https://doi.org/10.1142/S0217979295000148}{Int. J. Mod. Phys. B \textbf{9}, 295 (1995)}.
\bibitem{conditional_quantum} V. V. Shende, S. S. Bullock, and I. L. Markov, \href{https://doi.org/10.1109/TCAD.2005.855930}{IEEE Trans. Comput.-Aided Des. Integr. Circuits Syst. \textbf{25}, 1000 (2006)}.
\bibitem{semiclassical_algorithms} R. B. Griffiths and C.-S. Niu, \href{https://doi.org/10.1103/PhysRevLett.76.3228}{Phys. Rev. Lett. \textbf{76}, 3228 (1996)}.
\bibitem{MLMC} G. Carleo and M. Troyer, \href{https://doi.org/10.1126/science.aag2302}{Science \textbf{355}, 602 (2017)}.
\bibitem{MCMC} S. Chib and E. Greenberg, \href{https://doi.org/10.1080/00031305.1995.10476177}{Am. Stat. \textbf{49}, 327 (1995)}.
\bibitem{sign_problem} M. Troyer and U.-J. Wiese, \href{https://doi.org/10.1103/PhysRevLett.94.170201}{Phys. Rev. Lett. \textbf{94}, 170201 (2005)}.
\bibitem{DMC} M. H. Kalos, D. Levesque, and L. Verlet, \href{https://doi.org/10.1103/PhysRevA.9.2178}{Phys. Rev. A \textbf{9}, 2178 (1974)}.
\bibitem{PIMC} D. M. Ceperley, \href{https://doi.org/10.1103/RevModPhys.67.279}{Rev. Mod. Phys. \textbf{67}, 279 (1995)}.
\bibitem{constrained_DMC} D. M. Ceperley and B. J. Alder, \href{https://doi.org/10.1103/PhysRevLett.45.566}{Phys. Rev. Lett. \textbf{45}, 566 (1980)}.
\bibitem{contrained_PIMC} E. W. Brown, B. K. Clark, J. L. DuBois, and D. M. Ceperley, \href{https://doi.org/10.1103/PhysRevLett.110.146405}{Phys. Rev. Lett. \textbf{110}, 146405 (2013)}.
\bibitem{autoregressive_VMC} O. Sharir, Y. Levine, N. Wies, G. Carleo, and A. Shashua, \href{https://doi.org/10.1103/PhysRevLett.124.020503}{Phys. Rev. Lett. \textbf{124}, 020503 (2020)}.
\bibitem{quantum_Shannon_theory} M. M. Wilde, \textit{Quantum Information Theory, Second Edition} (Cambridge University Press, Cambridge, 2017).
\bibitem{entropy_bound} I. H. Kim, \href{https://arxiv.org/abs/1709.07828}{arXiv:1709.07828}.
\bibitem{autoregressive_classical} D. Wu, L. Wang, and P. Zhang, \href{https://doi.org/10.1103/PhysRevLett.122.080602}{Phys. Rev. Lett. \textbf{122}, 080602 (2019)}.
\bibitem{Bayesian_networks} J. Pearl, \href{https://doi.org/10.1016/0004-3702(86)90072-X}{Artif. Intell. \textbf{29}, 241 (1986)}.
\bibitem{Renyi_entanglement} M. B. Hastings, I. Gonz\'{a}lez, A. B. Kallin, and R. G. Melko, \href{https://doi.org/10.1103/PhysRevLett.104.157201}{Phys. Rev. Lett. \textbf{104}, 157201 (2010)}.
\bibitem{state_discrimination} A. Chefles, \href{https://doi.org/10.1080/00107510010002599}{Contemp. Phys. \textbf{41}, 401 (2000)}.
\bibitem{github_repo} The software implementation and the data and scripts used to generate figures are available at \url{https://github.com/godotalgorithm/ftvmc-test}.
\bibitem{barren_plateau} J. R. McClean, S. Boixo, V. N. Smelyanskiy, R. Babbush, and H. Neven, \href{https://doi.org/10.1038/s41467-018-07090-4}{Nat. Commun. \textbf{9}, 4812 (2018)}.
\bibitem{stochastic_gradients} A. Harju, B. Barbiellini, S. Siljam\"{a}ki, R. M. Nieminen, and G. Ortiz, \href{https://doi.org/10.1103/PhysRevLett.79.1173}{Phys. Rev. Lett. \textbf{79}, 1173 (1997)}.
\bibitem{transverse_1D_Ising} P. Pfeuty, \href{https://doi.org/10.1016/0003-4916(70)90270-8}{Ann. Phys. (N. Y.) \textbf{57}, 79 (1970)}.


\end{thebibliography}
\end{document}